\documentclass[twocolumn,twoside,slac_two]{revtex4}
\usepackage{graphicx}
\usepackage{fancyhdr}
\begin{document}

\title{Charm Mixing - Theory}

%

\author{E. Golowich}
\affiliation{Physics Department, University of 
Massachusetts, Amherst, MA 01003, USA}

\begin{abstract}
We discuss Standard Model (SM) 
and New Physics (NP) descriptions of D$^0$ mixing.  
The SM part of the discussion addresses both quark-level and hadron-level 
contributions.  The NP part describes our recent works on 
the rate difference $\Delta\Gamma_{\rm D}$ and the mass difference 
$\Delta M_{\rm D}$.  In particular, we describe how the recent 
experimental determination of $\Delta M_{\rm D}$ is found to place 
tightened restrictions on parameter spaces for 
$17$ of $21$ NP models considered in a recent paper by 
Hewett, Pakvasa, Petrov and myself.
\end{abstract}

\maketitle

\thispagestyle{fancy}


\section{Introduction}
Given the forthcoming operation of the LHC, 
perhaps the dominant role of experimental flavor studies 
in particle physics will be supplanted by discoveries 
in the so-called new physics.  Even if 
flavor physics faces an unsure future, all would 
acknowledge its remarkable recent progress via the observation of 
rare phenomena such as CP-violation in $B$-mesons or 
particle-antiparticle mixing for $B_s$ and $D^0$ mesons. 
If new physics is indeed observed, the continued exploration of 
rare observables could well be an asset in deciphering exotic LHC events.  

My purpose in this talk is to describe two recent theoretical 
contributions to $D^0$ 
mixing~\cite{Golowich:2006gq,Golowich:2007ka}.\footnote{
These two papers contain extensive references 
of contributions not possible to cite here.} 
Both Ref.~\cite{Golowich:2006gq} and Ref.~\cite{Golowich:2007ka} 
should be considered in the context of the recent 
HFAG values~\cite{peterson}, 
\begin{eqnarray}
& & x_{\rm D} \equiv {\Delta M_{\rm D} \over \Gamma_{\rm D}} = 
\left( 8.4^{+3.2}_{-3.4} \right) \cdot 10^{-3} 
  \nonumber \\
& & y_{\rm D} \equiv {\Delta \Gamma_{\rm D} \over 2 \Gamma_{\rm D}} =
\left( 6.9 \pm 2.1 \right) \cdot 10^{-3} \ \ .
\label{xy}
\end{eqnarray}
In light of the Physical Review Letters criteria of 'observation' 
($> 5\sigma$) or 'evidence' ($3\sigma$-to-$5\sigma$), we see that 
the above $2.4\sigma$ determination for $x_{\rm D}$ amounts to 
a 'measurement' ($< 3\sigma$).   As such, we all await improvments 
in sensitivity for charm mixing.  

The observed signal is seen to occur at about the $1\%$ level.  
Whether or not this is the magnitude expected for the SM signal 
is a topic I will discuss shortly.  At any rate, 
I wish to also consider the possibility of a NP component in 
$D^0$ mixing amplitude,
\begin{equation}
{\cal M}_{\rm mix} = {\cal M}_{\rm SM} + {\cal M}_{\rm NP} \ \ . 
\label{ampl}
\end{equation}
The relative phase between ${\cal M}_{\rm SM}$ 
and ${\cal M}_{\rm NP}$ is not known.  Thus, in our 
detailed study of various NP contributions to $x_{\rm D}$ 
in Ref.~\cite{Golowich:2007ka} we most often compared 
the NP predictions to $\pm 1\sigma, \pm 2\sigma$ windows 
relative to the central $x_{\rm D}$ value of Eq.~(\ref{xy}).

\subsection{Operator Product Expansion (OPE) and Renormalization 
Group}
An important technical aspect of 
Refs.~\cite{Golowich:2006gq, Golowich:2007ka} is the process of 
relating an amplitude at some NP scale $\mu = M$ to one at, say, 
the charm scale $\mu = m_c$.  This takes the form 
\begin{equation}
\label{SeriesOfOperators}
\langle f | {\cal H}_{NP} | i \rangle =
G \sum_{i=1} {\rm C}_i (\mu) ~
\langle f | {\cal Q}_i  | i \rangle (\mu) \ \ ,
\end{equation}
where the prefactor $G$ has the dimension of inverse-squared mass, the 
${\rm C}_i$ are dimensionless Wilson coefficients, and the 
${\cal Q}_i$ are the effective operators.  At the leading order 
of dimension six, it turns out that there are {\it eight} four-quark 
operators, 
\begin{eqnarray}
& & {\cal Q}_1 = (\overline{u}_L \gamma_\mu c_L) 
\ (\overline{u}_L \gamma^\mu
c_L)\ , \nonumber \\
& & {\cal Q}_2 = (\overline{u}_L \gamma_\mu c_L) \ (\overline{u}_R \gamma^\mu
c_R)\ , \nonumber \\
& & {\cal Q}_3 = (\overline{u}_L c_R) \ (\overline{u}_R c_L) \ , \nonumber \\
& & {\cal Q}_4 = (\overline{u}_R c_L) \ (\overline{u}_R c_L) \ , \nonumber \\
& & {\cal Q}_5 = (\overline{u}_R \sigma_{\mu\nu} c_L) \ ( \overline{u}_R
\sigma^{\mu\nu} c_L)\ , \nonumber \\
& & {\cal Q}_6 = (\overline{u}_R \gamma_\mu c_R) \ (\overline{u}_R \gamma^\mu
c_R)\ , \nonumber \\
& & {\cal Q}_7 = (\overline{u}_L c_R) \ (\overline{u}_L c_R) \ , \nonumber \\
& & {\cal Q}_8 = (\overline{u}_L \sigma_{\mu\nu} c_R) \ (\overline{u}_L
\sigma^{\mu\nu} c_R)\ \ .
\label{SetOfOperators}
\end{eqnarray}
Any given NP contribution will often involve several of 
these, but in all events never more than these eight.  
The evolution is determined by solving the RG equations obeyed by
the Wilson coefficients,
\begin{equation}
\label{AnomEq}
\frac{d}{d \log \mu} \vec C (\mu) = \hat \gamma^T \vec C (\mu)\ \ ,
\end{equation}
where $\hat \gamma$ is the $8\times 8$ anomalous dimension 
matrix~\cite{Ciuchini:1997bw}.  The output of this calculation 
is a set of RG factors $r_i (\mu, M)$ which are expressed in 
terms of ratios of QCD fine structure constants evaluated at 
different scales, {\it e.g.} as with 
\begin{eqnarray}
& & r_1(\mu,M) =  \\
& & \left(\frac{\alpha_s(M)}{\alpha_s(m_t)}\right)^{2/7}
\left(\frac{\alpha_s(m_t)}{\alpha_s(m_b)}\right)^{6/23}
\left(\frac{\alpha_s(m_b)}{\alpha_s(\mu)}\right)^{6/25} \ . \nonumber
\label{wilson}
\end{eqnarray}

\subsection{Operator Matrix Elements}
One needs ultimately to evaluate the $D^0$-to-${\bar D}^0$ matrix 
elements of the eight operators $\{ {\cal Q}_i \}$. In general, 
eight non-perturbative parameters would need 
to be evaluated by some means such as a lattice determination.  
As a practical matter, 
the method used in Refs.~\cite{Golowich:2006gq, Golowich:2007ka} 
is to introduce a `modified vacuum saturation' (MVS), where
all such matrix elements are written in terms of the known 
matrix elements of (V-A)$\times$(V-A) and (S-P)$\times$(S+P) matrix
elements $B_{\rm D}$ and $B_{\rm D}^{\rm (S)}$~\cite{Gupta:1996yt}, 
\begin{eqnarray}\label{ME_MVS}
& & \langle {\cal Q}_1 \rangle = {2 \over 3} f_{\rm D}^2 M_{\rm D}^2 B_D \ ,
\nonumber \\
& & \langle {\cal Q}_2 \rangle = -{1 \over 2} f_{\rm D}^2 M_{\rm D}^2 B_D
   - \displaystyle{1 \over N_c} f_{\rm D}^2 M_{\rm D}^2 
{\bar B}_{\rm D}^{\rm (S)} \ , \nonumber \\
& & \langle {\cal Q}_3 \rangle = \displaystyle{1 \over 4 N_c} 
f_{\rm D}^2 M_{\rm D}^2 B_D
   + {1 \over 2} f_{\rm D}^2 M_{\rm D}^2 {\bar B}_{\rm D}^{\rm (S)} 
\ , \nonumber \\
& & \langle {\cal Q}_4 \rangle = - \displaystyle{2 N_c - 1 \over 4 N_c} 
f_{\rm D}^2 M_{\rm D}^2 {\bar B}_{\rm D}^{\rm (S)} \ , \nonumber \\
& & \langle {\cal Q}_5 \rangle = \displaystyle{3 \over N_c} 
f_{\rm D}^2 M_{\rm D}^2 {\bar B}_{\rm D}^{\rm (S)} \ , \nonumber \\
& & \langle {\cal Q}_6 \rangle = \langle {\cal Q}_1 \rangle \ , \nonumber \\
& & \langle {\cal Q}_7 \rangle = \langle {\cal Q}_4 \rangle \ , \nonumber \\
& & \langle {\cal Q}_8 \rangle = \langle {\cal Q}_5 \rangle \ \ ,
\end{eqnarray}
where the number of colors is $N_c=3$ and, 
as in Ref.~\cite{Golowich:2005pt}, we define 
\begin{equation}\label{bbar}
{\bar B}_{\rm D}^{(S)} 
\equiv B_{\rm D}^{\rm (S)} \cdot {M_{\rm D}^2 \over (m_c+m_u)^2}  \ \ .
\end{equation}
With the above theoretical machinary in hand, we are now 
ready to consider SM and NP contributions to $D^0$ mixing.

\section{Standard Model Analysis}
One can use quarks or hadrons as the basic degrees of freedom 
in carrying out the SM analysis of $D^0$ mixing.  In principle, 
these should give the same result.  However, as we shall see, 
rather different features appear in each description.


\subsection{Quark-level Analysis} 
At leading order in the SM, the OPE for $D^0$ mixing consists of two 
dimension-six four-quark operators~\cite{Georgi:1992as}.  The next
order contains fifteen dimension-nine six-quark operators.  For each 
increasing order in the OPE, there are still more local quark and gluon 
operators and the problem of determining operator matrix elements 
becomes ever more severe.  For this reason, the dimension 
six sector has received by far the most attention.

\begin{figure}[t]
\includegraphics[width=14pc]{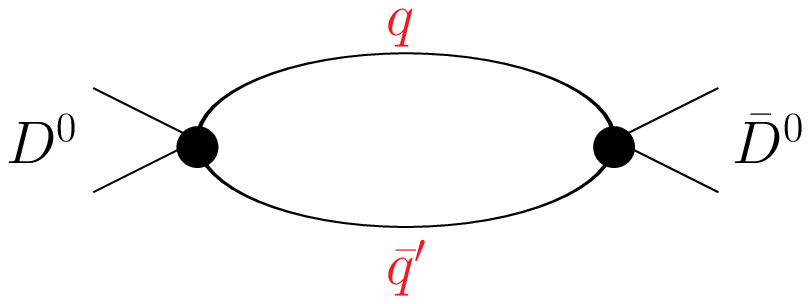}
\caption{Loop diagram for $D^0 \to {\overline D}^0$.}
\label{fig:loop}
\end{figure}

The dimension six amplitude is depicted in Fig.~\ref{fig:loop}.
Since the $b$-quark is essentially decoupled due to the tiny $V_{ub}$ 
value, only the light $d,s$ quarks propagate in the loop.
The Cabibbo dependence of this diagram, $\sin^2\theta_c$, itself 
seems to suggest that the experimental signal 
(near the $0.01$ level) is easily understood.
But not so fast!  For convenience, let us set $m_d = 0$.  Then the only 
mass ratio that appears in the problem is 
\begin{equation}
z\ \  \equiv \ \ (m_s/m_c)^2  \ \ \simeq \ \ 0.006 \ \ .
\label{mr}
\end{equation}
Table~\ref{tab:qkmix} examines one of the 
loop-functions for $\Delta\Gamma_{\rm D}$ and shows the results of
carrying out an expansion in powers of $z$.  We see that
the contributions of the individual intermediate states in the mixing 
diagram are {\it not} intrinsically small -- in fact, they begin to 
contribute at ${\cal O}(z^0)$.  
However, flavor cancellations remove all contributions through 
${\cal O}(z^2)$ for $\Delta\Gamma_{\rm D}$, 
so the net result is ${\cal O}(z^3)$.  
Charm mixing clearly experiences a remarkable GIM suppression! 

We understand the reason for this.  $D^0$ mixing vanishes in 
the limit of exact SU(3) flavor symmtery.  It is nonzero only 
because flavor SU(3) is broken, and indeed, $D^0$ mixing 
occurs at second order in SU(3) breaking~\cite{Falk:2001hx}.  
A factor of $z$ will accompany each order of SU(3) breaking 
and the rate difference $y_{\rm D}$ will experience 
an additional factor of $z$ due to helicity suppression.  

\begin{table}[h]
\caption{Flavor cancellations in $\Delta\Gamma_{\rm D}$.}
\begin{tabular}{cccc}
\hline\hline
Intermediate State & ${\cal O}(z^0)$ & ${\cal O}(z^1)$ & ${\cal O}(z^2)$\\
\hline
$s{\overline s}$ & $1/2$ & $-3z$ & $3z^2$ \\
$d{\overline d}$ & $1/2$ & $0$ & $0$ \\
$s{\overline d}+d{\overline s}$ & $-1$ & $3z$ & $-3z^2$ \\
\hline
Total & $0$ & $0$ & $0$ \\ \hline\hline
\end{tabular}\\[2pt]
\vskip .05in\noindent
\label{tab:qkmix}
\end{table}

Of course, this is just the leading order (LO) result 
in QCD, and we should consider the next-to-leading order 
result as well, 
\begin{eqnarray}
x_{\rm D} &=& x_{\rm D}^{({\rm LO})}+x_{\rm D}^{({\rm NLO})}\ \ ,
\nonumber \\
y_{\rm D} &=& y_{\rm D}^{({\rm LO})}+y_{\rm D}^{({\rm NLO})}\ \ .
\label{expan}
\end{eqnarray}
This has been done in Ref.~\cite{Golowich:2005pt} and the 
results are summarized in Table~\ref{tab:SD}, which 
reveals that $y_{\rm D}$ is given 
by $y_{\rm NLO}$ to a reasonable approximation (due to the
removal of helicity suppression by virtual gluons) 
whereas $x_{\rm D}$ is greatly 
affected by destructive interference between $x_{\rm LO}$ and 
$x_{\rm NLO}$.  The net effect is to render $y_{\rm D}$ and 
$x_{\rm D}$ of similar small 
magnitudes, at least through this order of analysis, as compared 
to the experiment signal.

\begin{table}[t]
\caption{Results at dimension-six in the OPE.}
\begin{tabular}{l||c||c|c} 
\colrule\hline 
 &LO & NLO & LO + NLO \\
\colrule \hline
$y_{\rm D}$ & $-(5.7 \to 9.5) \cdot 10^{-8}$
& $(3.9 \to 9.1)\cdot 10^{-7}$
& $\simeq 6 \cdot 10^{-7}$  \\
$x_{\rm D}$ & $-(1.4 \to 2.4) \cdot 10^{-6}$ 
& $(1.7 \to 3.0)\cdot 10^{-6}$ & 
$\simeq 6 \cdot 10^{-7}$ \\
\colrule \hline
\end{tabular}
\vskip .05in\noindent
\label{tab:SD}
\end{table}

It is not inconceivable that the quark-level prediction of 
$x_{\rm D}$ and $y_{\rm D}$ just described might be considerably 
affected by a higher order in the OPE~\cite{Ohl:1992sr} 
which suffers less $z$ suppression.  
Simple dimensional analysis~\cite{Bigi:2000wn} suggests the magnitudes 
$x_{\rm D} \sim y_{\rm D} \sim 10^{-3}$ might be achievable, although 
order-of-magnitude cancellations or enhancements are 
possible.

\subsection{Hadron-level Analysis} 
Most of the work involving the hadron degree-of-freedom has been 
done on $y_{\rm D}$.  One starts with the following general expression 
for $\Delta\Gamma_{\rm D}$, 
\begin{eqnarray}
& & \Delta\Gamma_{\rm D} = \frac{1}{M_{\rm D}}\, {\rm Im}\,I  
\label{delgam} \\
& & I \equiv \langle {\bar D}^0 |
    \,i\! \int\! {\rm d}^4 x\, T \Big\{
    {\cal H}^{|\Delta C|=1}_w (x)\, {\cal H}^{|\Delta C|=1}_w(0) \Big\}
    | D^0 \rangle \ . 
\nonumber 
\end{eqnarray}
To utilize this relation, one inserts intermediate states between the 
$|\Delta C| = 1$ weak hamiltonian densities ${\cal H}^{|\Delta C|=1}_w$.  
Although this can be done using either quark or hadron degrees of
freedom, let us consider the latter here.  Clearly, 
some knowledge of the matrix elements $\langle n |  
{\cal H}^{|\Delta C|=1}_w | D^0 \rangle$ is required.  

One approach is to model $|\Delta C| = 1$ decays theoretically and 
fit the various model parameters to charm decay data. 
Some time ago, $\Delta\Gamma_{\rm D}$ was determined 
in this manner and the result $y_{\rm D} \simeq 10^{-3}$ 
was found~\cite{Buccella:1994nf}.  This value is smaller than 
the recent BaBar and Belle central values.  

Alternatively, one can arrange for charm decay data to play a 
somewhat different role.  The earliest work in this regard focussed 
on the $P^+P^- = \pi^+\pi^-, K^+ K^-, K^- \pi^+, K^+ \pi^-$ 
states~\cite{Donoghue:1985hh,lw}.
In the flavor SU(3) limit, this subset of states gives zero 
contribution due to cancellations.  But SU(3) breaking had already 
been known to be significant in individual 
charm decays.   Since the study of charm decays 
in the 1980's lacked an abundance of data, these references could 
only conclude that '$y_{\rm D}$ might be large'.  

A modern version of this approach now exists, although the 
analysis takes an unexpected direction~\cite{Falk:2001hx}.  
Since SU(3) breaking occurs at second order in $D^0$ 
mixing, let us hypothesize that the contribution 
of the $P^+P^-$ sector is in fact negligible due to flavor cancellations.  
Likewise for all other sectors whose decays are kinematically
allowed.  However, this cannot be true for 
four-pseudoscalars because decay into four-kaon states is 
kinematically forbidden.  In Ref.~\cite{Falk:2001hx} it is 
estimated that these `kinematically-challenged' sectors  
can provide enough $SU(3)$ violation to induce 
$y_D\sim 10^{-2}$.  I personally find such an argument to be an 
important advance in our understanding of the subject.  
At the same time, it is unfortunately more persuasive than compelling 
due to the uncontrollable uncertainties inherent in this line of 
reasoning.  

To summarize, we have just described how the observed $D^0$ mixing signal 
could well arise from SM physics, but the associated numerical prediction is 
seen to be lacking in precision.  This conceivably leaves room for 
some NP mechanism to co-contribute or even dominate the SM signal.  
In the following we consider in turn NP analyses of the width
difference $y_{\rm D}$ and the mass difference $x_{\rm D}$.  

\section{NP and the Width Difference}
At first glance, it would appear unlikely that NP could affect
$y_{\rm D}$ because the particles contributing to the loop 
amplitude of Fig.~\ref{fig:loop} must be {\it on-shell}.  Since 
NP particles will be heavier than the charm mass, `there can be no 
NP contribution to $y_{\rm D}$'.  Or so goes the argument.  

However, as explained in Ref.~\cite{Golowich:2006gq}, NP effects in 
${\cal H}^{|\Delta C|=1}_w$ {\it can} generally contribute to 
$y_{\rm D}$.  In the loop amplitude of Fig.~\ref{fig:loop}, the NP 
contribution (empirically small for $\Delta C=-1$ processes)
arises from either of the two vertices.  
We represent the NP $\Delta C=-1$ hamiltonian as 
(indices $i,j,k,\ell$ represent color),
\begin{eqnarray}\label{HamNP}
&& {\cal H}^{\Delta C=-1}_{NP} =
\sum_{q,q'} \ D_{qq'}
\left[\overline {\cal C}_1(\mu) {\cal O}_1 +
\overline {\cal C}_2 (\mu) {\cal O}_2 \right]\ , \nonumber
\\
&& {\cal O}_1 = \overline{u}_i \overline\Gamma_1 q_j' ~
\overline{q}_j \overline\Gamma_2 c_i \ ,  \nonumber \\
&& {\cal O}_2 = \overline{u}_i \overline\Gamma_1 q_i' ~
\overline{q}_j \overline\Gamma_2 c_j\ , 
\end{eqnarray} 
where $D_{qq'}$ and the spin matrices $\overline\Gamma_{1,2}$ 
encode the NP model.  $\overline {\cal C}_{1,2}(\mu)$ are
Wilson coefficients evaluated at
energy scale $\mu$ and the flavor sums on $q,q'$ 
extend over the $d,s$ quarks. 

This leads to a prediction for the NP contribution to $y_{\rm D}$.  
For a generic NP interaction, one finds 
(with the number of colors $N_c = 3$) 
\begin{eqnarray}
\label{yNP}
& & y_{\rm D} \ = - \ \frac{4\sqrt{2} G_F}{M_{\rm D} \Gamma_{\rm D}} \
\sum_{q,q'} {\bf V}_{cq'}^* {\bf V}_{uq} D_{qq'}
\left(K_1 \delta_{ik}\delta_{j\ell} \right.
\nonumber \\
& & +
\left. K_2 \delta_{i\ell}\delta_{jk} \right)
\sum_{\alpha=1}^5 \ I_\alpha (x,x') \ \langle
\overline{D}^0| \ {\cal O}_\alpha^{ijk\ell} \ | D^0 \rangle, \ \
\end{eqnarray}
where $\{K_\alpha\}$ are combinations of Wilson coefficients, 
\begin{eqnarray}
& & K_1= \left({\cal C}_1 \overline{\cal C}_1 N_c
+ \left({\cal C}_1 \overline{\cal C}_2 + \overline{\cal C}_1
{\cal C}_2 \right)\right)\ \ , \nonumber \\
& & K_2 = {\cal C}_2 \overline{\cal C}_2 \ \  , 
\label{k}
\end{eqnarray}
and the $\{ {\cal O}_\alpha^{ijk\ell} \}$ are four-quark 
operators written down in Ref.~\cite{Golowich:2006gq}.  
Numerical results for some NP models are displayed in 
Table~\ref{tab:newy}.  

\begin{table}[h]
\begin{center}
\caption{Some NP Models and $y_{\rm D}$.}
\begin{tabular}{|l|c|c|}
\hline \textbf{Model} & \textbf{y$_{\rm D}$} & \textbf{Comment} 
\\
\hline RPV-SUSY & $6 \cdot 10^{-6}$ & Squark Exchange \\
 & -$4 \cdot 10^{26}$ & Slepton Exchange \\
\hline Left-right & -$5 \cdot 10^{-6}$ & `Manifest' \\
 & -$8.8 \cdot 10^{-5}$ & `Nonmanifest'  \\
\hline Multi-Higgs & $2 \cdot 10^{-10}$ & Charged Higgs \\
\hline Extra Quarks & $ 10^{-8}$ & Not Little Higgs \\
\hline
\end{tabular}
\label{tab:newy}
\end{center}
\end{table}

One sees that the entries, aside from R-parity violating SUSY, produce
small contributions.  We emphasize, however, that Eq.~(\ref{HamNP}) 
and Eq.~(\ref{yNP}) represent general formulae for the contribution
of all NP models of $|\Delta C|=1$ interactions, encompassing 
those not included in Table~\ref{tab:newy}.

\section{NP and the Mass Difference}
As the operation of the LHC looms near, the number of 
potentially viable NP models has never been greater. 
In this section, I will give an overview of 
Ref.~\cite{Golowich:2007ka}, whose hallmark is the study 
of {\it many} (21 in all) NP models.  Perhaps the best way to start is 
to consider the different ways that `extras' can be added to 
the SM: 
\begin{itemize}
\item Extra gauge bosons (LR models, {\it etc})
\item Extra scalars (multi-Higgs models, {\it etc})
\item Extra fermions (little Higgs models, {\it etc})
\item Extra dimensions (universal extra dims., {\it etc})
\item Extra global symmetries (SUSY, {\it etc})
\end{itemize}
Although this approach does not provide a totally clean 
partition of NP models ({\it e.g.} obviously SUSY contains 
extra particles appearing in other categories), it proved 
useful to the authors of Ref.~\cite{Golowich:2007ka}.

\begin{table}[t]
\caption{NP models studied in Ref.~\cite{Golowich:2007ka}}
\begin{tabular}{|c|}
\colrule\hline 
Model 
\\ \hline\hline
Fourth Generation  \ \ \\
$Q=-1/3$ Singlet Quark  \\
$Q=+2/3$ Singlet Quark  \\
Little Higgs  \\ 
Generic $Z'$ \\
Family Symmetries \\
Left-Right Symmetric  \\ 
Alternate Left-Right Symmetric \\ 
Vector Leptoquark Bosons \\
Flavor Conserving Two-Higgs-Doublet \\
Flavor Changing Neutral Higgs  \\
FC Neutral Higgs (Cheng-Sher ansatz) \\
Scalar Leptoquark Bosons  \\
Higgsless \\
Universal Extra Dimensions \\
Split Fermion  \\
Warped Geometries \\
Minimal Supersymmetric Standard \\
Supersymmetric Alignment \\
Supersymmetry with RPV \\
Split Supersymmetry \\ 
\hline\hline
\end{tabular}
\vskip .05in\noindent
\label{tab:bigtable}
\end{table}

The broad menu of NP models which were analyzed is listed in 
Table~\ref{tab:bigtable}.  The extensive content of this list 
({\it e.g.} there are {\it four} different SUSY realizations 
and {\it three} involving large extra dimension) indicates 
how rich the field of NP models has become.  Of course, the 
subject of NP is by now fairly mature (in preparing this talk, 
I realized that my first paper on charged Higgs 
bosons~\cite{Golowich:1978nh} was written 
nearly 30 years ago!) and thus many models  
have been well exposed to the scrutiny of experiment.  This would
seem to imply that parameter spaces for the various models have 
shrunk so much that a measurement like $D^0$ mixing would have 
little impact.  In fact, in giving this talk in several venues 
I challenged each audience to predict how many of the 21 models 
considered here were constrained by the $D^0$ mixing values or 
equivalently how many evaded constraint.  
Before answering this question, we consider a specific NP example 
in some detail.  

\begin{figure}[b]
\centerline{
\includegraphics[width=8cm,angle=0]{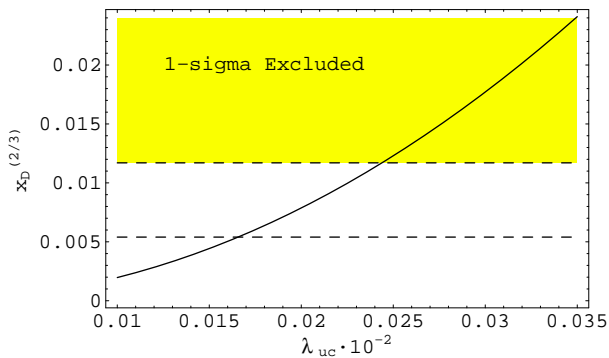}}
\vspace*{0.1cm}
\caption{Value of $x_{\rm D}$ as a function of the mixing parameter 
$\lambda_{uc}$ in units of
$10^{-2}$ in the $Q=+2/3$ quark singlet model.  The $1\sigma$ experimental
bounds are as indicated, with the yellow shaded area depicting the
region that is excluded.}
\label{SingletQuark23Fig}
\end{figure}

Suppose a {\it vector-like} quark of charge $Q=+2/3$~\cite{branco} 
is added to the SM.  
Recall that a vector-like quark is one whose electric charge 
is either $Q = + 2/3$ or $Q = - 1/3$ and which is an 
$SU(2)_{\rm L}$ singlet.  Both choices of charge are actually 
well motivated, as such fermions appear 
explicitly in several NP models. 
For example, weak isosinglets with $Q = -1/3$ appear in
$E_6$ GUTs~\cite{q2,jlhtgr}, with one for each of the three
generations ($D$, $S$, and $B$).  Weak isosinglets with $Q = +2/3$ 
occur in Little Higgs theories~\cite{q3a,q3b} in which the 
Standard Model Higgs boson is a pseudo-Goldstone boson, and the heavy
iso-singlet $T$ quark cancels the quadratic divergences generated
by the top quark in the mass of the Higgs boson. 
\begin{figure} [t]
\centerline{
\includegraphics[width=4cm,angle=0]{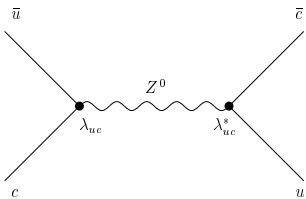}}
\caption{Tree-level contribution from $Z^0$-exchange.
\label{qfig2}}
\end{figure}
We restrict our attention here to the $Q = +2/3$ case. Since the 
electroweak quantum number assignments are different than those for the
SM fermions, flavor changing neutral current interactions 
will be generated
in the left-handed up-quark sector.  Thus, there will also be FCNC 
couplings with the $Z^0$ boson~\cite{branco}.  These couplings contain 
a mixing parameter $\lambda_{uc}$ which is constrained by the 
unitarity condition 
\begin{equation}
\lambda_{uc} \equiv - \left( V_{ud}^*V_{cd} + V_{us}^*V_{cs} + 
V_{ub}^*V_{cb} \right) \ \ .
\label{uc}
\end{equation} 
A tree-level contribution to $\Delta M_D$ is thus 
generated from $Z^0$-exchange (see Fig.~\ref{qfig2}).  It is 
straightforward to calculate that 
\begin{eqnarray}
x^{\rm (2/3)}_{\rm D} &=& {G_F \lambda^2_{uc} 
\over \sqrt 2 M_{\rm D} \Gamma_D}  r_1(m_c,M_Z) 
\langle \bar{D}^0| \ {\cal Q}_1 \ | D^0 \rangle
\nonumber \\
&=&  {2G_F f_{\rm D}^2  M_{\rm D} 
\over 3\sqrt 2\Gamma_D}  B_D  
\left( \lambda_{uc} \right)^2 r_1(m_c,M_Z)  
\end{eqnarray}
where we have made use of Eq.~(\ref{SetOfOperators}) and 
Eq.~(\ref{ME_MVS}).  The result is displayed in the graph of 
Fig.~\ref{SingletQuark23Fig}, which contrasts 
a $\pm 1\sigma$ window (dashed lines) about the HFAG 
central value with the NP prediction 
$x^{\rm (2/3)}_{\rm D}$ (solid line), as a function of 
the mixing parameter $\lambda_{uc}$.  The bound on 
$\lambda_{uc}$ from $D^0$ mixing turns out to be 
roughly two orders of magnitude better than that from the 
CKM unitarity constraint. 

Upon performing analogous analyses for the other NP models, 
we arrive at a set of constraints (mainly in the form of 
graphs) like the one depicted in Fig.~\ref{SingletQuark23Fig}.
It is not possible here to summarize the results for all 21 
models of Table~\ref{tab:bigtable}.  One must refer to 
Ref.~\cite{Golowich:2007ka} for that.  However, we can answer 
the question raised earlier about how many models avoid being 
meaningfully constrained by the $D^0$ mixing data.  The answer is just 
4 of the 21 models, which came as a surprise to many.
These 4 models are Split SUSY, Universal Extra Dimensions, 
LR Symmetric and Flavor-conserving Higgs Doublet.

It is of interest to briefly consider a few of these, in order 
to understand how the $D^0$ mixing constraints can be evaded:
\begin{enumerate}
\item Split SUSY~\cite{nimagian}: This is a relatively new variant 
of SUSY (2003-4) in which SUSY breaks at a very large scale 
$M_S \gg 1000$~TeV.  All scalars except the Higgs have mass
$M \sim M_S$ whereas fermions have the usual weak-scale mass.  
It is known that large $D^0$ mixing in SUSY will involve squark 
amplitudes.  But since squark masses in Split SUSY are huge, 
the mixing becomes suppressed.
\item Universal Extra Dimensions~\cite{ued}: UED is a variant of the 
idea that TeV$^{-1}$-sized extra dimensions exist.  There are 
no branes appearing in this approach, so all SM fields reside in the bulk 
and just one extra dimension is usually considered.  Each SM field 
will have an infinity of KK excitations. It turns out that 
GIM cancellations suppress all but a few b-quark KK terms, but these 
are CKM inhibited.  
\end{enumerate}  
So we see that supressions can arise from more than one source 
and that the suppressing mechanism will depend on the specific model.

\section{Conclusions}
At long last, signals for $x_{\rm D}$ and $y_{\rm D}$ have 
been observed.  These experimental findings, although greatly welcome, 
whet our appetite for ever more precise determinations.  
Hopefully these will be forthcoming, so we can put aside any 
lingering concerns that all the excitement has been the result of 
statistical fluctuations.

The SM analysis, as is so often the case, is not without its 
difficulties.  At the quark level, theoretical analysis 
in the dimension six sector through NLO gives $x_{\rm D} \sim 
y_{\rm D} \simeq 10^{-6}$.  These values are tiny compared to 
the reported experimental signals.  It is evident that the triple sum 
over the operator dimension $d_n$, the QCD coupling $\alpha_s$ 
and the mass expansion parameter $z$ of Eq.~(\ref{mr}) is 
slowly convergent.  This approach remains inconclusive at best.  

A more promising avenue is to study $y_{\rm D}$ with the hadronic 
degree of freedom.  This yields a plausible, and quite possibly correct, 
explanation for reaching the $y_{\rm D} \sim 0.01$ level.  Again, 
however, the effect of strong interaction uncertainties mars 
predictive power.

Finally, the work of Refs.~\cite{Golowich:2006gq, Golowich:2007ka} 
has explored which NP models can yield sizable values for 
$x_{\rm D}$, $y_{\rm D}$ and which cannot.  Charm mixing data 
has been found to infer useful constraints on NP parameters spaces, 
and as should be clear to all, provides a most welcome addition 
to the High Energy Physics community.

\begin{acknowledgments}
The author's contribution to the work described above was supported 
in part by the U.S.\ National Science Foundation under 
Grant PHY--0555304.
\end{acknowledgments}

\bigskip 

\end{document}